\documentclass[aps,prb,twocolumn,notitlepage,nofootinbib]{revtex4}
\usepackage[a4paper,top=2.00cm,bottom=2.00cm,left=2.00cm,right=2.00cm]{geometry}
\usepackage{amsfonts}
\usepackage{amscd}
\usepackage{bm}
\usepackage{amstext}
\usepackage{amsmath}
\usepackage{amssymb}
\usepackage{color}
\usepackage{enumerate}
\usepackage{graphicx}
\DeclareGraphicsExtensions{.eps, .pdf, .jpg, .tif}
\usepackage{hyperref}
\usepackage{float}
\graphicspath{{./}}
\setcitestyle{super}
\usepackage{mycolors}
\usepackage{bookmath}
\usepackage{mathfrak}
\usepackage{mymatrices}
\usepackage{mygroups}
\def\nicefrac#1#2{\genfrac{}{}{}{1}{#1}{#2}}
\begin{document}
\title{
Superfluidity in the Interiors of Neutron Stars
\\
\vspace{-20mm}
\hspace*{-10mm}{\tiny Lecture Notes for the NATO Advanced Study Institute on
{\sl Timing Neutron Stars},held in \c{C}e\c{s}me, Turkey, April 4-15, 1988.}
\newline
\hspace*{-50mm}{\tiny Published in NATO ASI Series C, Vol. 262, pp. 457-490, Kluwer Academic Press, 1989.}
\vspace*{10mm}
}
\author{J.A. Sauls} 
\email{sauls@lsu.edu}
\affiliation{Department of Physics \& Astronomy, Northwestern University, 
             Evanston, Illinois 60208, USA}
\begin{abstract}
The discoveries of more than 400 neutron stars as radio pulsars continue to provide an intellectual challenge to physicists and astronomers with diverse backgrounds. I review some of the ideas that have been proposed for the structure of neutron star interiors, and concentrate on the theoretical arguments for the existence of superfluidity in neutron stars. I also discuss the implications of neutron superfluidity and proton superconductivity for the rotational dynamics of pulsars, and review arguments that have been proposed for observable effects of superfluidity on the timing history of pulsars and perhaps other neutron stars.
\end{abstract}
\maketitle

\section{Introduction}

The discovery of pulsars, and their subsequent identification as rotating neutron stars, initiated a flurry of activity by theorists to better understand neutron-star structure and matter at extremely high density [see the reviews by Refs.\onlinecite{bay75a,bay79}]. Our model of neutron star structure is reminiscent of the interior structure of Earth;~\cite{and82} a solid outer crust, with various layers, which encloses a much hotter fluid core. This latter component is a conducting fluid, which although not well understood, is the source of a magnetic field. For the purpose of constructing a model for neutron star interiors, perhaps their distinguishing feature is that they are extraordinarily cold; even at interior temperatures of order $10^6\ -\ 10^8{\rm K}$ quantum statistics plays a crucial role in the thermodynamic and transport properties of nuclear matter. It is because neutron-star matter is cold that many exotic states of matter have been proposed to exist inside these stars. The idea that neutron stars contain a liquid interior of {\it superfluid neutrons} and {\it superconducting protons}~\cite{mig60,gin65} was motivated in large part by our understanding of the mechanism for superconductivity in terrestrial materials as a result of the BCS theory,~\cite{bar57} and also by the observation of {\it glitches} in the timing data of Vela pulsar.~\cite{rad69,rei69}

In Sec.~\ref{NS_Model} I review the {\it standard model} for the interior structure of neutron stars, paying special attention to the arguments in support of the idea that neutron stars contain superfluid interiors. One expects the rotational motion of a neutron star with a superfluid core, decelerating under the action of external radiation torques, to be rather different than an otherwise similar star with a normal fluid core of high viscosity. The main features of the rotational equilibrium of the superfluid and superconducting interior are discussed in Sections \ref{Superfluidity}-\ref{Rotation_Equilibrium_Protons}, while in the remaining sections I discuss the essential features of the rotational dynamics of the superfluid interior. The important differences in the rotational dynamics of a star with a superfluid core compared to a normal-matter core are:
(i) the timescales for momentum transfer between the superfluid and the neutron star crust, and (ii) the existence of metastable flow states which are fundamentally related to the phenomenon of persistent superfluid flow, as in liquid HeII, and vortex pinning, as in laboratory superconductors. It is here that an important connection exists between the theory of neutron stars and the timing observations on radio pulsars. In Sections~\ref{Glitches}-\ref{Open_Problems} I discuss more speculative aspects of the theory and some unanswered questions of importance for our understanding of the evolution of pulsar interiors. I begin by reviewing some concepts from the theory of superfluidity and superconductivity.

\section{Condensation}\label{Condensation}

Review articles [{\it e.g.} Ref.~\onlinecite{sha80}] that discuss superfluidity in neutron stars often emphasize the importance of the {\it energy gap} in the superfluid phase. The existence of an energy gap in nuclear matter is important in understanding neutron-star rotational dynamics; however, the essential concept is the phenomenon of {\it condensation}, by which I mean the {\it macroscopic occupation} of a single quantum state. In liquid $^4 He$ superfluidity is closely related to Bose-Einstein condensation. The relevant single-particle states are simply $\psi_{\vp} \sim e^{i\vp \cdot \vr}$, and below $2.2{\rm K}$ a finite fraction of the $^4 He$ particles occupy the zero-momentum state, $\psi$, \ie $|\psi|^2\sim \cO(N/V)$. The important feature of condensation, so far as the phenomenon of superfluidity is concerned, is that the amplitude of the condensate,
\be
\psi (\vR ) = |\psi (\vR )| \ e^{i\vartheta (\vR )}
\,,
\ee
is {\it phase coherent} over the entire fluid. Thus, if the condensate phase is known at point $\vR$, then one can predict the phase a macroscopic distance away, according to $\vartheta ({\vR}^{'} )= \vartheta (\vR ) + 2M{\vv}_s \cdot ({\vR}^{'} - \vR )/\hbar$, where ${\vv}_s$ is the local velocity of the condensate, \ie the superfluid velocity.

In systems of Fermions, \eg neutrons and protons in the interior of neutron stars, condensation occurs by the formation of {\it pairs} of Fermions, or {\it Cooper pairs}.  Since Fermions have a spin ($s=\hbar /2$ for neutrons and protons) the amplitude of the condensate depends on the internal arrangement of the constituent spins; in addition the pair may exhibit internal orbital motion. The general form of the Cooper pair amplitude is described by a wave function $\psi_{s_1 ,s_2}(\vR ,\vr)$, where $s_1 ,s_2$ are the spin projections of the Fermions, $\vR$ is the center-of-mass of the pair, and $\vr$ is the orbital coordinate of the pair. The dimension of the pair wave function in neutron matter, the {\it orbital size} of the Cooper pair, is of order $100 {\rm fm}$, which although small, is nevertheless large compared to the average distance between neutrons in the interior of the star. Even though the size of the pair wave function is measured in hundreds of Fermis, this amplitude is {\it coherent} over macroscopic distances, in this case throughout the liquid interior of the star.  When condensation occurs a macroscopic number of neutrons form pairs in precisely the same two-particle wave function, independent of their center-of-mass position. Hereafter I use the term `order parameter' to mean `Cooper pair amplitude' because this macroscopically occupied state represents a high degree of order, and the symmetry and structure of the Cooper pair amplitude determine the macroscopic magnetic and flow properties of the condensed phase.  There is a great variety of phenomena associated with the spin and orbital motion of the Cooper pairs.  Since these states may play a role in the theory of the rotational motion of neutron star interiors, it is useful to classify some of the possible internal motions of the pairs and comment briefly on what is known about the order parameters for laboratory superfluids and superconductors.

\subsection{S-wave, spin-singlet pairs}

Since the order parameter represents a bound state of two Fermions it must be anti-symmetric under exchange of the coordinates and spins of the pair,
\be
\psi_{s_1,s_2}(\vR,\vr)=- \psi_{s_2,s_1}(\vR,-\vr)
\,.
\ee
Most laboratory superconductors are described by an order parameter with quantum numbers, $|\vS|=0$ ({\it spin singlet}) and $|\vL|=0$ ({\it s-wave}), where $\vS=\vs_1 + \vs_2$ is the total spin, and $\vL=\vr\times\vp$ is the orbital angular momentum of the pair. The orbital motion is isotropic and the spins of the Fermions are paired into a magnetically inert singlet; thus, the pair amplitude reduces to a single complex scalar amplitude, $\psi(\vR) = \psi_{\uparrow,\downarrow}$, analogous to the order parameter in superfluid $^4 He$. This is also the form of the order parameter believed to describe the condensate of superfluid neutrons in the inner crust of a neutron star, and the superconducting protons in the liquid interior.


\subsection{P-wave, spin-triplet pairs}

The most remarkable terrestrial superfluids are the phases of liquid $^3He$.~\cite{and78} There are three superfluid phases that are stable in different regions of temperature, pressure and magnetic field. This fact alone differentiates liquid $^3He$ from liquid $^4 He$ and conventional {\it s-wave} superconductors. It is known that superfluid $^3 He$ is described by a {\it spin-triplet} ($S = |{\vs}_1 + {\vs}_2 | = \hbar $), {\it p-wave} ($L=\hbar$) order parameter. For pairing into states with one unit of orbital angular momentum, $\psi$ is a linear combination of the spherical harmonics, $\{ Y_{1,m}(\vr )\ ;\ m= \pm 1 ,0 \}$,
\be
\psi_{s_1,s_2}(\vR,\vr)=\sum_{m=\pm 1,0} \psi_{s_1,s_2}^m(\vR)\,Y_{1,m}(\vr)
\,.
\ee
These odd-parity states, $[\ Y_{1,m}(\vr )= -Y_{1,m}(-\vr )\ ]$, imply that the spin-dependent part of the pair amplitude is symmetric under exchange of the two Fermion spins. The $^3 He$ atom has a total spin of $\hbar /2$ due to an unpaired nucleon, and there are three symmetric spin states that can be constructed from two spin-$1/2$ amplitudes. Thus, the general form of the pair amplitude is
\be
\ket{\psi}=\psi_{\uparrow\uparrow}\,\ket{\uparrow\uparrow}
          +\psi_{\uparrow\downarrow}\,\ket{\uparrow\downarrow+\downarrow\uparrow}
	  +\psi_{\downarrow\downarrow}\,\ket{\downarrow\downarrow}
\,.
\ee
In contrast to superfluid $^4 He$, a spin-triplet p-wave superfluid such 
as $^3 He$ requires up to nine complex amplitudes (3 spin $\times$ 3 orbital).
Note that these Cooper pairs are in principle {\it magnetic}. I
list the form of the order parameter for a few specific cases.

\subsection{Superfluid $^3He-B$}

The order parameter is a superposition of all three magnetic states with
equal amplitudes and phases,
\be
\hspace*{-2mm}
\ket{\psi}\ns=\ns\psi_B\{Y_{1,-1}\ket{\uparrow\uparrow} 
                           + Y_{1,0}\ket{\uparrow\downarrow\ns+\ns\downarrow\uparrow}
	                   + Y_{1,1}\ket{\downarrow\downarrow}\}
\,,
\ee
and the orbital amplitude is such that the {\it total} angular momentum of the Cooper pairs is zero, $|\vJ | = | \vL + \vS | = 0$. The B-phase is a special state which is ``isotropic'' in that the pair amplitude is invariant under joint rotations of the spin and orbital coordinates.\footnote{This statement is slightly modified when the weak nuclear dipolar interaction is included.}

\subsection{Superfluid $^3He-A_1$}

The $A_1$ phase corresponds to pairing in only one component of the spin triplet and is stable only in a magnetic field and a narrow range of temperatures. The order parameter directly reflects the {\it magnetic polarization} of this superfluid,
\be
\ket{\psi} =\psi_{A_1}\,Y_{1,1}\,\ket{\uparrow\uparrow}
\,.
\ee

\subsection{Interior superfluid of neutron stars: $^3P_2$ phase}

As I discuss below it is plausible that neutron matter in the liquid interior of a neutron star is a Fermion superfluid described by a spin-triplet, p-wave amplitude with total angular momentum $J=2$,~\cite{hof70}
\be
\ket{\psi} = \sum_{J_z = 0,\pm 1,\pm 2}\,\psi_{J_z}\,\ket{J=2, J_z} 
\,.
\ee
In fact the ground state of the non-rotating $^3P_2$ phase, to use the spectroscopic designation for the pair amplitude, is believed to be a state with $J_z=0$ with respect to a fixed but arbitrary axis $\vz$.~\cite{sau78,vul84} Thus, the ground state of the core superfluid in neutron star matter is also described by a single scalar amplitude, $\psi_0$. This is no longer the case for the equilibrium state of a {\it rotating} neutron superfluid; a proper description of the {\it vortices} in the $^3P_2$ phase - which are required for the superfluid to co-rotate with the crust, conducting plasma and magnetic field - requires that all five magnetic sub-states, $\psi_{J_z}$, be present in the vicinity of the vortices. This fact leads to a novel magnetic structure for the vortex lines inside neutron stars (c.f.  Sec.~\ref{Vortices_3P2}).

\section{Pairing Instability and Transition Temperatures}\label{NS_Model}

There is no direct evidence that the interiors of neutron stars are superfluid. However, there are two arguments in favor of this idea. The first is based on the BCS theory of superconductivity, which is arguably the most successful many-body theory of condensed matter. The second reason is the existence of long timescales for the recovery of the angular deceleration of several pulsars following a glitch (c.f. Sec.~\ref{Mutual_Friction}).

\begin{figure}
\centerline{
\includegraphics[width=\columnwidth]{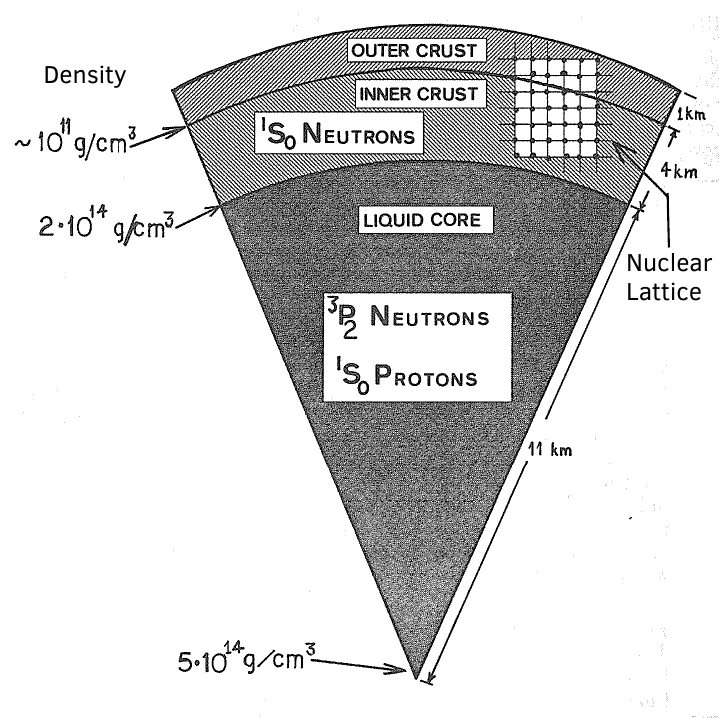}
}
\caption{{\bf Standard Model} - Structure of a Neutron Star.}
\label{fig1}
\end{figure}

The basic structure of a neutron star with a mass $M=1.4\ M_{\rm sun}$ is summarized in Fig.~\ref{fig1}. The radius and central density of the star, which depend on the mass and the equation of state of neutron-rich nuclear matter for densities above that of terrestrial nuclear matter, are both somewhat uncertain. However, all models of neutron stars have a liquid interior, which contains most of the moment of inertia of the star, surrounded by a solid metallic crust of neutron-rich nuclei embedded in a degenerate fluid of electrons. The radial structure of the crust has been studied in detail by numerous authors and is reviewed by Ref.~\onlinecite{bay75a}. Of particular importance is the structure of the inner crust of the neutron star for densities $\rho>4.3\times 10^{11}{\rm g/cm}^3$, where the nuclei become so neutron rich that the neutrons begin leaking out of the nuclei to form a background fluid of degenerate neutrons surrounding the nuclear lattice. This crustal region persists to densities near terrestrial nuclear matter density, $\rho\simeq 2\times 10^{14}{\rm g/cm}^3$, at which point the nuclei dissolve into a dense fluid consisting primarily of neutrons and a small percentage of protons and electrons, all of which are degenerate. Many other exotic states of matter have been proposed to exist in very dense cores of neutron stars, including pion condensates, free quarks and solid neutron matter.~\cite{bay75a} However, I do not discuss these more speculative possibilities for the inner core.

\begin{figure}
\centerline{
\includegraphics[width=\columnwidth]{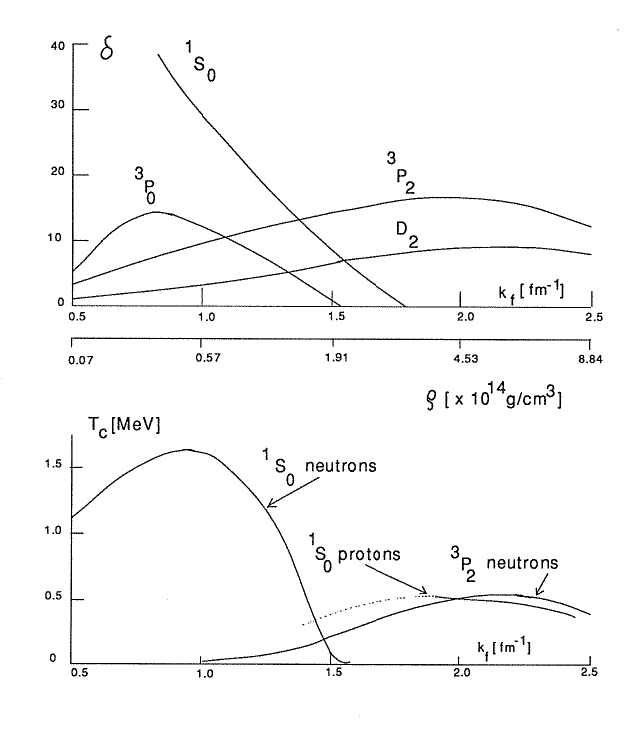}
}
\caption{Nucleon-nucleon phase shifts and $T_c$ vs. density.}
\label{fig2}
\end{figure}

Neutron stars are cold ({\it i.e.} $T\sim 10^8{\rm K}\ll T_{\rm Fermi}\sim 10^{12}{\rm K}$) and the same theoretical arguments that lead to the conclusions that terrestrial matter should be superconducting with transition temperatures $T_c \simeq 10^{-3} \ T_{\rm Fermi}$ also predict that neutron stars should have superfluid interiors. The necessary ingredient for the formation of a condensate of Cooper pairs, and hence a superfluid (or superconductor) is an attractive interaction between two neutrons (or protons) on the Fermi surface with {\it zero} total momentum. The Fermi sea guarantees the formation of a bound-state, \ie a Cooper pair, no matter how weak the interaction, so long as it is attractive. Of course the strength of the interaction has an important effect on the temperature at which condensation occurs. In neutron-star matter the origin of this attraction is the nucleon-nucleon interaction, which has the contributions,
\be
V_{nn} = V_{\rm central}(|\vr|) + V_{\rm so}(|\vr|)\,\vS\cdot\vL
\,,
\ee
where the {\it central} part of the potential is attractive at long-range, $r > \onehalf{\rm fm}$, due the exchange of pions, and repulsive at short distances due to the exchange of the $\omega$ meson; this same vector meson is responsible for the spin-orbit interaction, which is large at short distances.~\cite{brown76} A great deal is known about these basic interactions from nucleon-nucleon scattering. In Fig. \ref{fig2} I reproduce the experimentally determined scattering phase shifts of free neutrons [as compiled in Ref.~\onlinecite{tam70}]. A positive phase shift represents an attractive interaction in channels with various angular momentum quantum numbers $(S,L,J)$. The energy dependence (in the center-of-mass frame) is converted to density by setting the center-of-mass energy equal to that of two Fermions on the Fermi surface, $\ie\,E_{\rm cm}=4E_F(\rho)$. At low density, below approximately $2\times 10^{14}{\rm g/cm}^3$, the most attractive channel is the singlet, s-wave channel ($^1S_0$). However, the $^3P_2$ and $^1D_0$ interactions dominate the S-wave interaction at higher density, with the $^3P_2$ channel being the most attractive. Note that the P-wave interactions with $J=0,1$ are always repulsive at high density. Based on this data, and calculations of the structure and density profile of a neutron star, Hoffberg,\et argued that more than one superfluid state was possible inside a neutron star.~\cite{hof70} In the inner crust, $3\times 10^{11}{\rm g/cm}^3<\rho<2\times 10^{14}{\rm g/cm}^3$, a BCS-superfluid of neutron pairs in $^1S_0$ bound states forms, while at higher densities the neutrons condense into a $^3P_2$ state. The lower density protons are predicted to condense into a $^1S_0$ state.~\cite{cha72} Many authors have used this phase shift data, combined with more sophisticated approaches, to estimate the transition temperatures for condensation into these superfluid states. Typical values of the transition temperatures, $T_c$, for both superfluid states range from $0.1{\rm Mev}$ to $1{\rm Mev}$, \ie $10^{9}{\rm K}$ to $10^{10}{\rm K}$, which are low temperatures compared to the Fermi temperatures of neutron-star matter, but quite high temperatures compared to the ambient temperatures for even the youngest neutron star; \eg the interior temperature of the Crab pulsar is estimated to be of order $10^8{\rm K}$.~\cite{alp85}

A word of caution: transition temperatures are notoriously difficult to calculate accurately. This is clear from the BCS formula for the transition temperature, $T_c = E_F\ e^{1/N(E_F)V_{BCS}}$, which contains in the exponent the strength of the pairing interaction, which itself is a many-body effective interaction between neutron excitations and may differ significantly from the bare interaction. The uncertainty in estimates of $T_c$ is in fact more serious than indicated by this simple formula. The BCS theory is an inadequate theory for predicting whether a given material will be a superconductor, \ie in predicting $T_c$. Such a theory exists, and was formulated roughly ten years after the BCS theory, but it is applicable only to superconductors in which the pairing interaction between electrons is mediated by the phonons of the (heavier) ionic lattice [for a review see Ref.~\onlinecite{rai86}]. There is so far no reliable first principles theory of $T_c$ for a self-interacting Fermi superfluid.\footnote{There is a rather lengthy literature on failed attempts to calculate the transition temperature and pairing channel for the superfluid phases liquid $^3He$ before it was discovered.} However, the {\it standard model} illustrated in Fig. \ref{fig1} is based on plausible estimates for the pairing channel and transition temperatures, probably the best estimates available given the current state of the art in many-body theory. A better model of neutron star structure will necessarily have to wait until a first-principles theory of the superfluid transition temperature in a self-interacting system is developed. Nevertheless, the {\it discovery} of superfluidity in liquid $^3 He$ gives us confidence in the BCS pairing theory as a mechanism for superfluidity in neutron stars, simply because the mechanism for pairing in neutron stars is the self-interaction between the nucleons. Before the discovery of superfluidity in liquid $^3 He$ it had not been demonstrated that superfluidity could arise from the self-interaction between the Fermions.

Although it is difficult to reliably predict $T_c$ for neutron-star matter, it is important to note that the BCS theory is an excellent theory {\it if $T_c$ and the pairing channel [$S,L,J$] are known.} It has the power to reliably predict

\begin{itemize}
\item[$\bullet$] the ground-state order parameter $\psi$,
\item[$\bullet$] thermodynamic and transport properties of the superfluid phase,
\item[$\bullet$] the hydrodynamic properties of {\it rotating} superfluids, and
\item[$\bullet$] the structure of vortices, an important consideration
                 for rotating P-wave superfluids.
\end{itemize}

\noindent Extensions of the BCS theory are sufficiently powerful that difficult problems of relevance to the rotational dynamics of superfluid neutron stars are also tractable, including,

\begin{itemize}
\item[$\bullet$] theoretical estimates of the pinning energies of vortex lines
                 on impurities or defects in the stellar crust,
\item[$\bullet$] theory of nucleation and destruction of vorticity at interfaces, 
                 \eg the crust-liquid interface, and
\item[$\bullet$] theoretical analysis of the mechanisms and timescales for dissipative
                 motion of vortex flow during deceleration or acceleration 
		 events of pulsars.
\end{itemize}

Below I review some aspects of the theory of superfluidity as it applies to a rotating neutron star, discuss some of the novel features of the {\it mixture} of core superfluids, and present a mechanism for rapid equilibration of the interior superfluid to a disturbance of the crustal rotation period.

\section{Superfluidity, Currents, and Quantized Circulation}\label{Superfluidity}

I assume for simplicity that the interiors of neutron stars are described by a scalar order parameter, \ie a $^1S_0$ pair amplitude. This is consistent with the standard model for the neutron liquid in the inner crust and the proton superconductor in the core, but not for the neutrons in the core. However, most of the concepts discussed here for the $^1S_0$ superfluid are easily generalized to the $^3P_2$ superfluid in the interior. In Secs.~\ref{Vortices_3P2} and~\ref{Glitches} I discuss the important differences between the $^3P_2$ and $^1S_0$ phases that reside in the core and crust, respectively.

The order parameter for the $^1S_0$ superfluid is described by an amplitude and a {\it phase},
\be
\psi(\vR)=|\psi|\,e^{i\vartheta(\vR)}
\,,
\ee
which have two distinct roles. The amplitude $|\psi|$ is a thermodynamic variable of state, fixed by a free-energy functional which attains its minimum in equilibrium. The free-energy is derivable from the BCS theory, and is most conveniently discussed in the limit $T \sim T_c$, the Ginzburg-Landau (GL) limit where the amplitude $|\psi|$ may be assumed small.~\cite{gin50} The GL free-energy functional is an expansion of the full BCS free-energy functional in terms of the order parameter $\psi$,
\be\label{GL_Free_Energy}
F[\psi]=\int_{\rm V} d^3R \Big\{\alpha|\,\psi|^2 
+
\nicefrac{1}{2}\beta\,|\psi|^4 
+
\nicefrac{1}{2}K\,|\grad\psi|^2\Big\}
\,.
\ee
The form of the GL expansion is required by gauge and rotational invariance of the free energy. 
The coefficients are obtained from the BCS theory: $\alpha=\bar\alpha(T/T_c-1)$, $\bar\alpha=N_F$, $\beta=(7\zeta(3)/8)$~$N_F/(\pi T_c)^2$, $K=N_F(7\zeta(3)/12)(\hbar v_F/2\pi\kb T_c)^2$, where $\zeta(3)\approx 1.2020569\ldots$, $N_F=3n/4E_F$ is the density of states for degenerate neutron matter at the Fermi energy, $n=k_F^3/3\pi^2$ is the neutron density expressed in terms of the Fermi wavevector, $k_F$, $v_F$ is the Fermi velocity, and $T_c$ is the superfluid transition temperature.~\cite{sau84}

In a uniform system the gradient term may be neglected, in which case the minimum of the functional is either the {\it normal state} with $\psi_{\rm eq} = 0$ for $T > T_c$, or the {\it condensed state} with $\psi_{\rm eq}^2 = \frac{\bar\alpha}{\beta}(1-T/T_c)$, and $F_{\rm eq}(T) = F[\psi_{\rm eq}]=-{\rm V}\,[\frac{\bar\alpha^2}{2\beta}](1-T/T_c)^2$ for $T<T_c$ is the condensation energy associated with pair formation.

The gradient energy in Eq.(\ref{GL_Free_Energy}) is related to the {\it kinetic energy} of superfluid flow. The connection between superflow and the phase of the order parameter is obtained by considering the transformation property of the order parameter under a Galilean {\it boost}.~\cite{mer78} The order parameter represents a bound-state of Cooper pairs, so we require that $\psi$ transform as a two-particle wave function, $\psi\xrightarrow[]{\vu}\psi\,e^{-i 2M\vu\cdot\vR/\hbar}$, where $\vu$ is the boost velocity and $M$ is the bare mass of the Fermion. Thus, the quantity,
\be\label{velocity_n}
\vv_s\equiv{{\hbar}\over {2M}}\grad\vartheta
\,,
\ee
transforms as a velocity field under a Galilean boost. That ${\vv}_s$ implies the existence of a mass current, $\vg$, is also evident from the transformation of the free-energy functional, $F\xrightarrow[]{\vu}F - \int d^3 R\,\{\vg\cdot\vu + \cO(u^2)\}$ under a Galilean boost.
Thus, the mass current density is proportional to the superfluid velocity: $\vg=\rho_s{\vv}_s,$ with a density $\rho_s\propto|\psi|^2$. This result defines the mass current in the rest frame of the {\it excitations}, \ie the non-condensate fraction with density, $\rho_{\rm ex}=\rho-\rho_s$, where $\rho$ is the total mass density of the fluid.

Many of the hydrodynamic properties of superfluids and superconductors follow directly from the form of the superfluid velocity field. Since ${\vv}_s$ is the gradient of a scalar field, superflow is \emph{irrotational}, 
%
\be
\grad\times\vv_s = 0
\,,
\ee
{\it except at singular points within the fluid.} This qualification is of crucial importance in the rotating state of a superfluid; the global circulation is given by the integral of $\vv_s$ around a {\it path C} that encloses the fluid,
\be
\oint_C \vv_s\cdot d\vl\,=\,{h\over{2M}}\,N
\,,
\ee
where $N$ is an integer. The right side of this equation is determined by the requirement that the order parameter be single-valued, equivalently that the phase change, $\Delta \vartheta_C$, around the path $C$ be an integral multiple of $2\pi$. This quantization of the circulation leads immediately to the concept of quantized vorticity and the the requirement that quantized vortices be present in a rotating vessel of superfluid.~\cite{ons49,fey55} In particular if $N\not= 0$ then there is necessarily a singularity in the velocity field. For a rectilinear line singularity with $N=1$, enclosed by a circular path of radius $R$, we have by inspection,
\be\label{vortex_flow}
\vv_s={\kappa\,{\hat{\vphi}}\over {2\pi R}}
\,,
\ee
which is the axial flow field of a vortex with a unit of circulation, $\kappa={h\over {2M}}$, and a singular vorticity field, $\grad\times\vv_s=\kappa\,\delta^{(2)}(\vR)\,\hat\vz$.

\section{Rotating Equilibrium of the Core of a Neutron Star}\label{Rotation_Equilibrium_Neutrons}

Thermodynamic equilibrium of a rotating vessel - in this case the crust and magnetic field of the neutron star - is determined by the free-energy functional {\it in the rotating frame}; only in this reference frame is the interaction between the particles of the liquid and the vessel time independent. The general form of this free energy is
\be F' = F - \vOmega\cdot\vL
\,,
\ee
where $F$ is the free-energy functional in the non-rotating frame, $\vL$ is the angular momentum of the fluid, and $\vOmega$ is the angular velocity of the vessel, \ie the crust of the neutron star.

\begin{figure}
\centerline{
\includegraphics[width=\columnwidth]{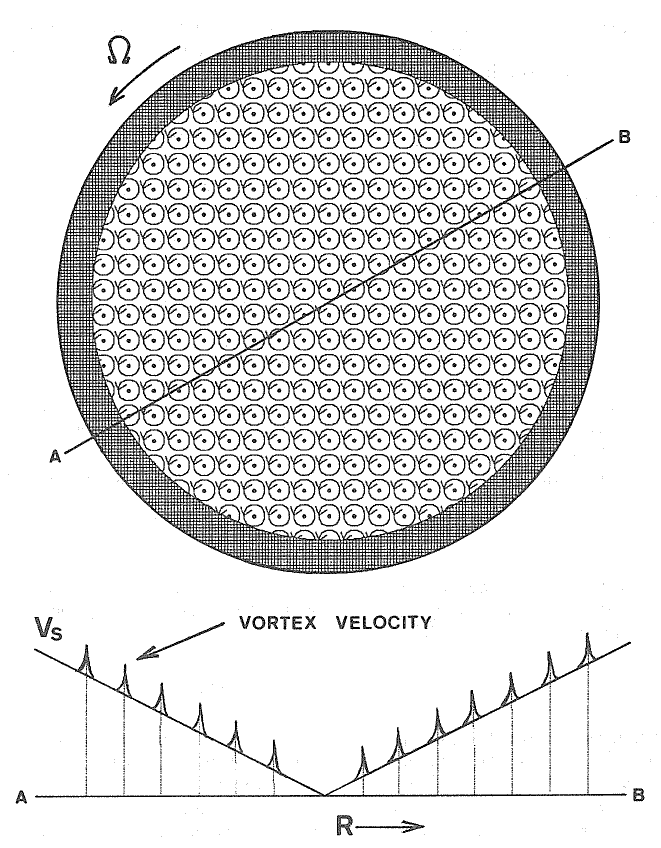}
}
\caption{The vortex state of a rotating neutron star.}
\label{fig3}
\end{figure}

This functional for the neutron superfluid simplifies in the limit where the order parameter $\psi_n$ is determined by its local equilibrium value, \ie $\psi_n=\psi_{\rm eq}\,e^{i\vartheta_n(\vR)}$, which is an excellent approximation in those cases in which one is interested in the {\it macroscopic} flow state of the fluid. However, the assumption of local equilibrium of the condensate breaks down on short length scales near the singularity of a phase, but for now it is sufficient to ignore this issue. I also ignore for the moment the fact that the protons are most likely superconducting. The angular momentum then reduces to the two-fluid form,
\be
\vL\,=\,\int d^3 R\,\vR\times(\rho_{s_n}\vv_{s_n}+\rho_{\rm ex}\vv_{\rm ex})
\,,
\ee
where $\vv_{s_n}=\nicefrac{\hbar}{2M_n}\grad\vartheta_n$ is the neutron superfluid velocity and $\rho_{s_n}=K\,(2M_n/\hbar)^2\,|\psi_{\rm eq}|^2$ is the neutron superfluid density.
The free-energy in the rotating frame reduces to
\be
F_n^{'}=F_n+\int d^3 R\,\onehalf\rho_{s_n}(\vv_{s_n}-\vOmega\times\vR)^2
\,,
\ee
where $F_n$ is independent of $\vv_{s_n}$ and $\vOmega\times\vR=\vv_{\rm ex}$ is the velocity of the normal-fluid excitations that are co-rotating with the crust and plasma interior. 
An unconstrained minimization of this functional leads to the incorrect conclusion that the superfluid co-rotates perfectly with the crust, \ie $\vv_{s_n}=\vOmega\times\vR$, in conflict with the constraint $\vv_{s_n}=\nicefrac{\hbar}{2M_n}\grad\vartheta_n$. In order for the superfluid to carry circulation, and thus to rotate with the vessel, the condensate must be perforated with vortices, each with a unit of circulation $\kappa_n=h/2M_n$, whose total circulation adds up to the rigid-body circulation of $2\Omega$. This latter condition is obtained by averaging the superfluid velocity over an area that contains many vortices, in which case the circulation contained in an area $\pi R^2$ of radius $R$ is
\be
\overline{\oint_{C_R}\vv_{s_n}\cdot d\vl}=2\Omega\times\pi R^2=N_v\,\nicefrac{h}{2M_n}
\,,
\ee
which yields the Onsager-Feynman formula for the areal density of vortices,~\cite{ons49,fey55}
\be\label{Onsager-Feynman}
\frac{N_v}{\pi R^2}=\frac{4M_n\Omega}{h} 
\simeq 6.3\times 10^3\,\frac{{\rm vortices}}{{\rm cm}^2}\,P^{-1}
\,,
\ee
where $P$ is the period of rotation of the star in seconds. For Vela pulsar ($P=0.083\,[{\rm sec}]$) this corresponds to an inter-vortex distance of approximately $4\times 10^{-3}\,{\rm cm}$. In Fig.~\ref{fig3} a sketch of the equilibrium rotating state of the core superfluid is shown, as well as that of the superfluid velocity along a line through the center of rotation.\footnote{A more accurate representation of the vortex state would show the vortices arranged in a hexagonal array with the area per vortex given in Eq.(\ref{Onsager-Feynman}).}
The superfluid velocity deviates from the classical rigid-body value of $\Omega R$ near the center of a vortex, where the velocity field of that particular vortex dominates the average field of all other vortices. In Sections \ref{Vortex_Core_Structure}-\ref{Electron_Vortex_Scattering} I discuss the {\it core} structure of vortices and their specific role in the rotational dynamics of the superfluid.

An important feature to note from Eq.(\ref{Onsager-Feynman}) is that the number of vortices is directly proportional to the angular speed of the crust. Thus, if the neutron star experiences a torque which decelerates the crust to lower speed, then a new equilibrium state can be achieved only by the {\it destruction} of vortices. This process proceeds by the {\it outward flow} of vortices, and annihilation of vorticity at the interface between the superfluid and the crust. Since the neutron stars that have been observed are rotating with speeds ranging from roughly $1-10^3\,{\rm rad/sec}$, and decelerating due to radiation torques acting on the magnetic field and crust of the star, a question of central importance for understanding the dynamics of a decelerating superfluid neutron star is: {\it what determines the timescale for the equilibration of the vortex density to the rotational speed of the crust?} The answer is that there is a {\it mutual friction force} between vortices and the non-superfluid component of the star. However, before discussing mutual friction, and the resulting deceleration of the superfluid component of the star, it is worthwhile to discuss the equilibrium rotation of the {\it superconducting} proton condensate.

\section{Rotational Equilibrium of the Superconducting Protons}
\label{Rotation_Equilibrium_Protons}

The important difference between the equilibrium state of the rotating superconductor (protons) and that of the neutral superfluid is that the superconducting condensate co-rotates with the crust lattice {\it without forming vortices}. This fact, first noted by F. London, follows from the hydrodynamic free-energy for the rotating superconductor, which has a similar form to that of the neutral superfluid,~\cite{london50}
\be
F_p' = F_p + \int d^3 R\,
     \Big\{\onehalf\rho_{s_p}\left(\vv_{s_p}-\vOmega\times\vR\right)^2 
     + \frac{|\vb|^2}{8\pi}\Big\}
\,,
\ee
where $\rho_{s_p}$ and $\vv_{s_p}$ are the proton condensate density and velocity, respectively.
In the case of charged supercurrents the free energy must necessarily include the self-consistently determined  magnetic field, $\vb=\grad\times\vA$, where $\vA$ is the vector potential. I temporarily omit the interaction between the neutron and proton condensates until Sec.~\ref{Superfluid_Drag}.
For the proton condensate the superfluid velocity field is given by
\be\label{velocity_p}
\vv_{s_p} = \frac{\hbar}{2M_p}\grad\vartheta_p - \frac{e}{M_p c}\vA(\vR)
\,,
\ee
where the appearance of the vector potential $\vA$ is required for local gauge invariance of the theory. Minimization of the free energy in the rotating frame again implies that the proton condensate velocity co-rotates with the crust of the neutron star. And in contrast to the neutral superfluid there is {\it no constraint} on the proton superfluid velocity field that is in conflict with the condition of co-rotation. In fact co-rotation of the bulk of the superconducting condensate is enforced with $\grad\vartheta_p=0$,~\footnote{For clarity I omit the stellar field, which generates proton vortex lines but does not change the substance of this argument.}
\be
\therefore\,\,\grad\times\vv_{p}= 2\,\vOmega = -{e\over{M_p c}}(\grad\times\vA)
\,.
\ee
Thus, the kinetic energy of the superconductor is minimized at the cost of a tiny magnetic field (the London field) distributed uniformly throughout the superconducting interior of the star,
\be
\vb_{\rm London}=-{{2M_p c}\over{e}}\vOmega
\,,
\ee
and directed along the axis of rotation. The source of this field is a thin boundary layer (of order $100\,{\rm fm}$ thick) of superconducting protons slightly out of co-rotation with the crust. Thus, co-rotation of the superconducting component with the crust and degenerate electron fluid is achieved at the cost of a tiny field (of order $10^{-4}$ Gauss for Vela), compared to stellar fields of order $10^{12}$ Gauss.

\section{Mutual Friction - Coupling of the Core Superfluid to the Crust}
\label{Mutual_Friction}

\begin{figure}
\centerline{
\includegraphics[width=\columnwidth]{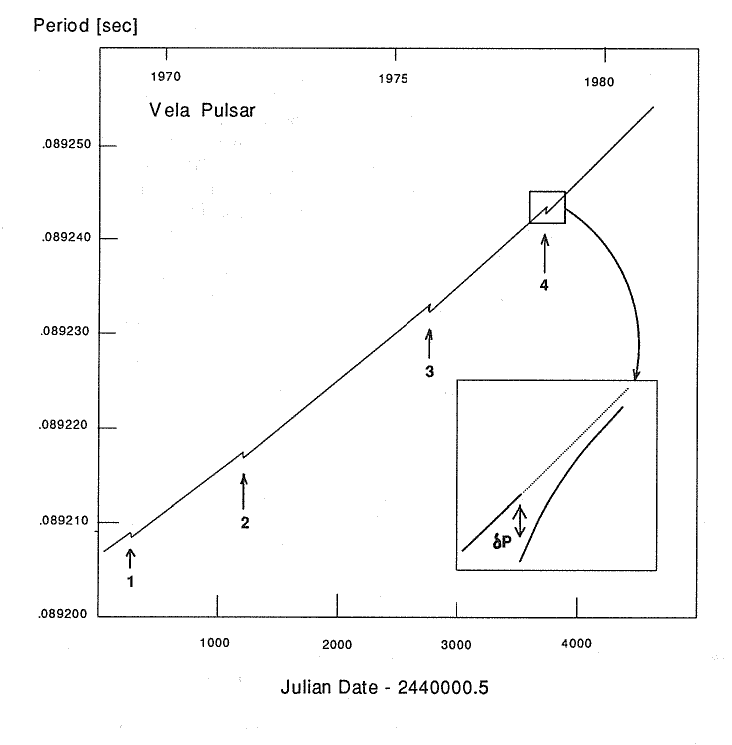}
}
\caption{Timing data of Vela pulsar showing the first four glitch events.}
\label{fig4}
\end{figure}

Figure \ref{fig4} shows a portion of the timing data for Vela pulsar, including the first four {\it glitches} [see Ref.~\onlinecite{dow81} for original references]. These glitches are discontinuous spin-up events ($\Delta \Omega \sim 10^{-6}\ \Omega$) of the neutron star, at least within the resolution of several days, accompanied by a discontinuous increase in the angular deceleration $\Delta\dot{\Omega}\sim -10^{-2}$. Following each of these glitches is a slow recovery of the angular deceleration back to the pre-glitch spin-down rate. The timescale for the recovery of the glitch is {\it a macroscopic timescale}, of order a few months or longer in the Vela. To date there have been seven giant glitches of the Vela pulsar occurring every 2 to 4 years since the timing observations began in 1969. The Crab pulsar also shows glitches, a total of 3 glitches of smaller magnitude $\Delta\Omega\sim 10^{-8}\ \Omega$, and the timescale for recovery of glitches in Crab varies from 3 to 60 days, also a macroscopic timescale. Glitch events have been observed in less studied pulsars, and seem to be ubiquitous, at least among relatively young pulsars. The largest glitch observed was in PSR 0355 + 54 with a magnitude of $\Delta\Omega=4.4\times 10^{-6}\,\Omega$.~\cite{lyn87}

As a means of defining the {\it mutual friction} timescale governing the coupling of the neutron superfluid interior to the rotation of the crust, I review the phenomenological two-component model for the rotational dynamics of a neutron star.~\cite{bay69b} This model supposes that the relevant structure of a neutron star is a {\it crust},\footnote{I use the term `crust' to refer to the solid outer crust, magnetic field and plasma interior of the star together, unless it is necessary to specify the individual constituent.} with moment of inertia $I_c$, containing a liquid interior of moment of inertia $I_s$. These two components are presumed weakly coupled via a {\it frictional coupling} of the form,
\be
N_{\rm internal}=I_s\dot{\Omega}_s = \left(\frac{I_c}{I}\right)[\Omega_c-\Omega_s]/\tau
\,,
\ee
which acts to bring the crust (rotating at $\Omega_c$) and interior fluid (rotating at $\Omega_s$) into co-rotation. The quantity $\tau$ that defines this coupling is the mutual friction timescale and $I=I_c+I_s$ is the total moment of inertia. The equation determining the rotational motion of the crust is
\be
I_c\dot{\Omega}_c = N_{\rm external} - N_{\rm internal}
\,.
\ee
Implicit in the model is the assumption that the relaxation of the fluid occurs nearly uniformly throughout the interior. Such a bulk mechanism for the coupling seems reasonable given that the interior fluid contains a high conductivity plasma of electrons and protons, which are strongly coupled to the the stellar magnetic field, and therefore the crust. 
The two-component model was proposed in order to explain the response of a neutron star to a glitch, and although it fails to explain the rotational history of the Vela or Crab pulsar quantitatively, it is the link to the tenuous thread of evidence supporting the proposal that neutron stars contain superfluid interiors. To appreciate this point it is important to examine the possible mechanisms for momentum transfer between the neutral liquid interior and the crust.

Easson has previously analyzed the coupling of the high conductivity plasma of protons and electrons to the magnetic field and crust with a simplified model in which the plasma is confined to a slab that is bounded on both sides by a conductor (``the crust'').~\cite{eas79} The plasma and conductor extend to infinity in the radial direction and a magnetic field $\vB$, perpendicular to the slab, penetrates the plasma and conductor. Easson analyzes the solutions to the magnetohydrodynamic equations with the initial condition that the rotation of the `crust' changes by $\Delta \Omega_c$. Spin-up of the plasma proceeds either by the formation of an Ekman boundary layer, and an associated radial flow of plasma which transports angular momentum, or by the excitation of low frequency hydromagnetic waves. In either case the spin-up time for the plasma is of order a few seconds for typical neutron-star parameters: $\tau_{\rm Ekman}\simeq 30\,T_7\,\Omega_2^{-\tiny 1/2}\,R_6\,\rho_{13}^{-\tiny 7/12}$ sec, and $\Omega_2 = \Omega /(10^2 {\rm rad/sec)}$, $R_6 = R/10^6 {\rm cm}$, $\rho_{13} = \rho / (10^{13} {\rm g/cm}^3)$, and $B_{12} = B/(10^{12} {\rm G)}$. Thus, for the purposes of analyzing the post-glitch response of a neutron star the plasma can be assumed to co-rotate with the solid crust during a glitch. The long timescale for the post-glitch relaxation observed in pulsars is then attributed to the equilibration of the neutral component of the star to the plasma and crust.\footnote{The model of Ref.~\onlinecite{eas79} assumes the proton matter is not superconducting. The spin-up of a type II superconductor, in which the field is organized into flux tubes has not been analyzed. However, Ref.~\onlinecite{alp84} provides some qualitative arguments for the rapid spin-up of the superconducting protons.}

The primary {\it bulk} scattering mechanism available for the transfer of momentum between the plasma and the neutron liquid interior is the strong interaction. It is straight-forward to estimate the timescale for momentum transfer between the neutron liquid and proton component of the plasma for the degenerate Fermi liquid of {\it non-superfluid} neutrons and {\it non-superconducting} protons. The timescale is determined primarily by the phase-space for binary collisions between a dilute {\it gas} of neutron and proton excitations at temperature $T \ll E_F$:~\cite{pines66}
\be
{\hbar \over {\tau_{np}}} \sim E_F\ \Big({T\over {E_F}}\Big)^2
\,,
\ee
which corresponds to a microscopic timescale, $\tau_{np} \sim 10^{-11}$ sec at $T= 10^6{\rm K}$. It is because this process leads to rapid equilibration that superfluidity is introduced. In order for the neutrons to be weakly coupled to the crust it is necessary that this strong-interaction scattering process be shut off, that the bulk of the neutron and proton excitations be {\it frozen out of the star}. This is most easily accomplished if there is an {\it energy gap}, $\Delta_n \gg T$, below which there are no allowed neutron states. Just such an energy gap appears as a consequence of the BCS pairing theory. In fact the timescale for momentum transfer at interior temperatures of $T \sim 10^6{\rm K}$, when the neutrons and protons are both superfluid (with $\Delta_n \sim \Delta_p \sim 1\,MeV$), becomes incredibly long,
\be
{\hbar \over {\tau_{np}}} \sim E_F \ e^{-(\Delta_n + \Delta_p )/T}
\Longrightarrow \tau_{np} \rightarrow \mbox{``}\infty\mbox{''}
\,,
\ee
far too long to account for the observed post-glitch timescales ranging from weeks to months. Thus, there is necessarily another mechanism responsible for the frictional coupling between the crust and the neutral interior.

\section{Vortex Structure and Electron-Vortex-Excitation Scattering}
\label{Vortex_Core_Structure}

\begin{figure}
\centerline{
\includegraphics[width=\columnwidth]{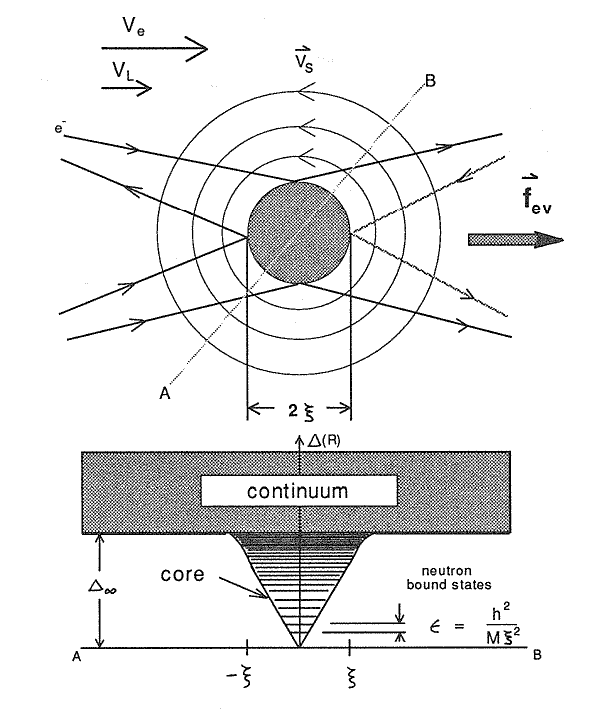}
}
\caption{Vortex structure for an s-wave vortex and electron-vortex scattering.}
\label{fig5}
\end{figure}

I previously represented a vortex in the s-wave neutron superfluid by the velocity field given in Eq.(\ref{vortex_flow}), with the {\it amplitude} of the order parameter given by the equilibrium amplitude, $|\psi(\vR)|=\psi_{\rm eq}$; the full order parameter for the vortex being
\be
\psi(\vR)=\psi_{\rm eq}\,e^{i\phi}
\,,
\ee
where the phase, $\phi$, is the azimuthal angle in coordinates centered on the vortex. This representation of the vortex is valid only on length scales long compared to the superfluid coherence length, $\xi$, defined roughly as the distance from the center of the vortex at which the superfluid kinetic energy density, $\tinyonehalf\,\rho_n{v_n}^2$, becomes equal to the condensation energy density, $\tinyonehalf\,\bar\alpha\,(1-T/T_c) {\psi_{\rm eq}}^2$. For neutron matter this length scale (for $T \ll T_{cn}$) is of order,
\be
\xi\simeq\,{{\hbar v_{Fn}}\over {\pi\Delta_n}}\simeq 10^2\,\mbox{fm}
\,,
\ee
and defines the radial dimension of the vortex core, inside of which the amplitude collapses to zero. The core is important for the rotational dynamics of the neutron star because it is the point of contact between the conducting plasma in the interior of the star and the neutron matter; at temperatures well below the neutron gap, $T \ll \Delta_n$, all scattering mechanisms involving neutron excitations in the bulk of the interior are frozen out. Scattering of the conducting plasma off the neutral component occurs only in the vicinity of the vortex cores. In fact since the protons are expected to be superconducting, while the electrons are not,\footnote{The superconducting transition for electrons, due to the polarization of the protons, is exceedingly small, $T_{ce} \simeq T_{Fe}\,e^{-1/\lambda}$ with $\lambda \sim {{e^2}\over{\hbar c}}\sim 1/137$.~\cite{bay75}} the only significant scattering processes are those involving the neutron vortex cores and the {\it electronic} component of the plasma. A schematic representation of the momentum transfer process between the electrons and the neutron vortices is shown in Fig. \ref{fig5}. The relative velocity between the electron fluid and the vortices, produced for example by a glitch, leads to preferential scattering of electrons from the vortex cores.

The equation of motion for rectilinear vortices moving relative to the background of excitations, in this case the electronic fluid, is well known from the study of superfluid hydrodynamics in liquid helium [see the review by Ref.~\onlinecite{son87}]. The momentum transfer to the vortex due to scattering of excitations off the core determines the response of the {\it superfluid} according to
\be
\vf_{ev}\ns=\ns -\rho_{\rm plasma}\ {{({\vv}_l - {\vv}_e )}\over
{n_v \tau}} = {{\hbar}\over {2M_n}} \rho_n ({\vv}_l - {\vv}_n)
\times \hat{\vOmega}
\,,
\ee
where ${\vv}_l$ is the velocity of the vortex line, ${\vv}_e$ is the velocity of the electrons, ${\vv}_n$ is the velocity of the neutron superfluid, $n_v$ is the areal density of the vortices and $\tau$ is the velocity relaxation time for the relative motion of the vortices and the electrons. This mutual friction timescale has been calculated for several models of the coupling of the plasma to the neutron vortices,~\cite{fei71,sau82a,alp84} and is simply related to the timescale for the dynamical response of the superfluid neutrons to a change in the motion of the plasma,~\cite{alp88} 
$\tau_d=\tau\,(\rho_s/\rho_{\rm plasma})(\nicefrac{n_v\kappa}{2\Omega})\sim\tau/x$, where $x\sim 0.05$ is the electron concentration in the interior.

The obvious mechanism of momentum transfer in the interior superfluid is the scattering of electrons, via electromagnetic interactions, off the {\it low-energy} neutrons that are bound to the vortex core. That such neutron bound states exist in the vicinity of the core is plausible given that the order parameter, and therefore the local gap, is depressed in the center of the vortex core (c.f. Fig.~\ref{fig5}). Even though the neutron gap vanishes inside the core, the lowest energy neutron state is determined by the dimensions of the vortex core; the spatially varying gap acts as a potential for the neutron excitations, and a simple estimate of the energy level spacing for bound states gives,
\be
\epsilon = {h^2 \over {M_n \xi^2}} \sim {{\Delta_n}^2
\over {E_F}} \ \ll \Delta_n
\,.
\ee
This level spacing determines the probability for a thermally excited neutron excitation in the vortex core,
\be
P_{\rm excitation}\sim\,e^{-{{\Delta_n^2}\over{E_F\,T}}}
\,,
\ee
which although much larger than that for bulk neutron excitations, is still an extremely small number, except in very young neutron stars. The density of excitations is the most sensitive factor determining the scattering rate for electrons interacting via their magnetic moments with these neutron excitations in the vortex cores. Feibelman's calculation of the scattering rate yields the estimate,~\cite{fei71}
\be
\tau\propto\frac{\Delta_n}{T}\,e^{{{\Delta_n^2}\over{E_F\T}}}\sim\,10^{20}\,{\rm sec}
\,,
\ee
for $\Delta_n=1\,{\rm Mev}$ at $T=10^6{\rm K}$. In all models of neutron stars, except those with high interior temperatures and low neutron gaps ($\ie \Delta_n^2/E_F T\sim 1$), electron-vortex-excitation scattering is ineffective, and probably does not explain the observed relaxation timescale following a glitch. The scattering time is so sensitive to the gap and interior temperature that is is difficult to account for the relatively small range of post-glitch relaxation times in pulsars with widely different ages, and presumably different interior temperatures. In any event there is a more efficient mechanism for momentum transfer in the interior that is {\it not} sensitive to the interior temperature and neutron gap.

\section{Vortices in the $^3P_2$ Neutron Superfluid}
\label{Vortices_3P2}

\begin{figure}
\centerline{
\includegraphics[width=\columnwidth]{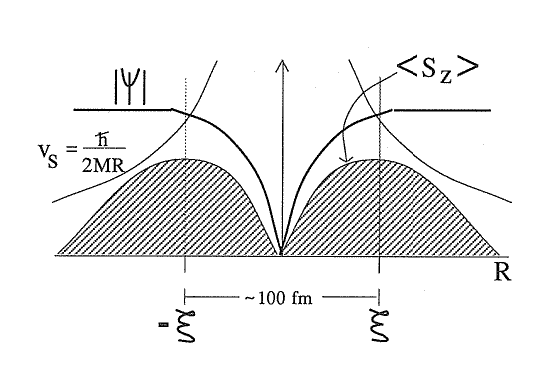}
}
\caption{Structure of a $^3P_2$ vortex showing the spin polarization.}
\label{fig6}
\end{figure}

I have so far treated the neutron superfluid interior as if the condensate were simply an s-wave, singlet state described by a single complex order parameter. This simplification is adequate for a description of the hydrodynamic flow far from the core of a vortex, but fails dramatically at distances of order the coherence length near any vortex in the $^3P_2$ phase. The qualitatively new feature of vortices in the $^3P_2$ phase is that the {\it condensate} in the core of the vortex is {\it spin-polarized} [For a more general discussion of vortex states in the $^3P_2$ phase see Refs.~\onlinecite{sau80,muz80,ric72}.],
\be
\langle S_z \rangle 
= 
|\psi_{\uparrow\uparrow}(R)|^2 - |\psi_{\downarrow\downarrow}(R)|^2
\,.
\ee
This can only occur in a {\it spin-triplet} superfluid, and since the neutrons have a magnetic moment the vortex itself carries a magnetization of order,
\be\label{Vortex_Magnetism}
M_{\rm vortex}\simeq(\gamma_n\,\hbar)\,n_n\,
\left({{\Delta_n}\over {E_F}}\right)^2\,\simeq 10^{11}\,{\rm G}
\,.
\ee
A sketch of the vortex structure is shown in Fig. \ref{fig6}. Magnetic vortices in a neutral superfluid were first proposed for the $^3P_2$ phase of neutron matter,~\cite{sau80,sau82a} but have since been observed experimentally in the B-phase of rotating $^3 He$, with the magnetization predicted by Eq.(\ref{Vortex_Magnetism}) (of course with the appropriate parameters for $^3 He$). The experimental observation of this effect in superfluid $^3 He$ gives us considerably more confidence in applying the microscopic theory to the novel phases of superfluid neutron-star matter.

The existence of a magnetic field localized near each neutron vortex is important for the rotational dynamics of the neutron superfluid because this inhomogeneous field scatters electrons. This mechanism for the transfer of momentum between the plasma and the neutron vortices is intrinsically different than Feibelman's mechanism because the vortex magnetization is a property of the {\it condensate} rather than the excitations. As a result the mutual friction timescale does not depend on the small number of thermally excited neutrons in the vortex cores, and is therefore only weakly dependent on the temperature and density,~\cite{sau82a}
\be
\tau=1.26 \times 10^8 \frac{k_{F_n} x^{2/3} P}{\Delta_n}\,{\rm sec}
\,,
\ee
where $k_{F_n}$ is the neutron Fermi wavevector in $fm$, $\Delta_n$ is the neutron gap in ${\rm MeV}$, $x$ is the electron concentration, and $P$ is the rotation period in seconds. For Vela pulsar this result gives a velocity relaxation time of about $\tau\sim 2\,{\rm months}$ with typical estimates of the gap and interior density, which is in reasonable agreement with the observed times for Vela. 
However, this agreement is destroyed by a more efficient scattering mechanism due to the {\it strong interaction} between the neutron and proton condensates (which is distinct from the strong interaction scattering between neutron and proton excitations). Below I show how the rapid equilibration of the core superfluid comes about and then in Sec.~\ref{Glitches} return to the question on the origin of the {\it slow} relaxation timescale observed in pulsars.

\section{Neutron-Proton Interactions and Superfluid Drag}\label{Superfluid_Drag}

\begin{figure}
\centerline{
\includegraphics[width=0.7\columnwidth]{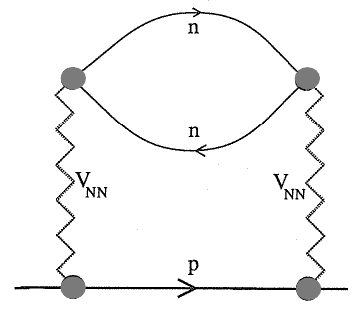}
}
\caption{Contribution to the self-energy of the proton from the strong interaction with the neutron Fermiliquid.}
\label{fig7}
\end{figure}

There is a larger magnetic field attached to each neutron vortex, which is independent of the spin structure of the order parameter, and leads to a rapid equilibration of the interior superfluid to the plasma with $\tau\simeq 400\,P\,{\rm sec}$! Most discussions of the hydrodynamics of neutron star interiors treat the constituents as independent fluids of electrons, protons, and neutrons, at most coupled together by electromagnetic interactions and the strong stellar field. In fact there is an important role played by the strong interaction between the neutrons and protons in the superfluid hydrodynamics of the interior fluid mixture that is distinct from the scattering of neutron and proton excitations.\footnote{I have previously mentioned that the {\it scattering} between neutron and proton excitations in the bulk is essentially irrelevant.} In a system of interacting Fermions, the elementary excitations are not simply {\it bare} neutrons or protons, but rather are quasiparticles - bare neutrons (or protons) {\it dressed} by a polarization cloud of other particles. This polarization cloud is a well-studied many body effect, and is responsible for the {\it effective mass} of a neutron (or proton) quasiparticle. In an interacting mixture of neutrons and protons the polarization cloud comprises {\it both neutrons and protons}. Calculations of the neutron and proton effective masses in neutron-star matter have been carried out by several authors (a contribution to the proton effective mass from polarization of the neutron medium is shown in Fig.~\ref{fig7}). In particular, Ref.~\onlinecite{sjo76} has shown that the neutron and proton effective masses, defined as the ratios of their respective Fermi momenta to their Fermi velocities, are given by
\ber
M_n^* =& M_n + \delta M_{nn}^* + \delta M_{np}^*\,,
\nonumber
\\
M_p^* =& M_p + \delta M_{pp}^* + \delta M_{pn}^*\,,
\eer
where $\delta M_{np}^*$ ($\delta M_{pn}^*$) determines the proton (neutron) contribution to the effective mass of the neutron (proton), and $\frac{\delta M^*_{pn}/M_p}{\delta M^*_{np}/M_n}=\frac{M^*_p n_n}{M^*_n n_p}$. The dilute concentration of protons interact with the neutrons through the long-range attractive part of the nucleon-nucleon interaction and reduce the neutron effective mass. Estimates of the neutron correction to the proton effective mass are $\delta M_{pn}^*\sim 0.5\,M_p$.

The polarization cloud that surrounds a neutron excitation in the two-component mixture of neutrons and protons is modified by the condensation of both the neutrons and protons, and as a result the superfluid mass current of neutrons is also modified; the constitutive relations are,
\ber
\vg_n &=& \rho_{nn}\ \vv_n + \rho_{np}\ \vv_p
\,,
\nonumber
\\
\vg_p &=& \rho_{pp}\ \vv_p + \rho_{pn}\ \vv_n
\,,
\eer
where the densities, $\rho_{nn},\rho_{pp}$ and $\rho_{np}=\rho_{pn}$ determine the {\it conserved} neutron and proton currents, $\vg_n$ and $\vg_p$, in terms of the superfluid velocity fields, $\vv_n$ and $\vv_p$, given in Eqs.~\eqref{velocity_n} and \eqref{velocity_p}. These equations, first considered by Andreev and Bashkin for \He-\Hefour\ mixtures~\cite{and75a} [see also Refs.~\onlinecite{var81,alp84}], exhibit the {\it superfluid drag effect} in which the condensate velocity of one species, \eg the neutrons, induces a particle current of the other species, \eg the protons. This effect is important because the rotation of the star couples directly to the velocity, $\vv_n$, which as I argued earlier is non-zero due to the existence of vortices in the neutron condensate. In the reference frame of the rotating star the proton condensate rotates with the crust without the formation of proton vortices, and thus the only contribution to the proton condensate velocity is the London current, $\vv_p = -\nicefrac{e}{M_p c}\vA$, which gives zero contribution to the bulk proton current in the rotating frame except near a neutron vortex line. The resulting superfluid charge current, induced by the neutron vortex lines, is
\be
\vj={c\over{4\pi}}(\grad\times\vb)={e\over{M_p}}[\rho_{pp}
\vv_p+\rho_{np}\vv_n]
\,,
\ee
where $\vb = \grad\times A$ is the induced magnetic field. Attached to each neutron vortex is a magnetic flux line with a local magnetic field determined by,
\be
\nabla^2 \vb + \Lambda_*^{-2}\vb = {{4\pi e}\over{M_p c}}\rho_{np}\,\grad\times\vv_n
\,,
\ee
where the vortex circulation, $\grad\times\vv_n=\kappa_n\delta^{(2)}(\vR)\,\hat\vz$, is the source of the flux and $\Lambda_*=\left[\nicefrac{M_p c^2}{4\pi e^2\rho_{pp}}\right]^{\tinyonehalf}$ is the length scale on which the magnetic field decays away from the center of the vortex. A simple calculation gives the magnitude of the trapped flux,
\be
\Phi_* = \oint \vA \cdot d\vl = \Phi_0 \Bigr({{M_p}\over{M_n}}\Bigl)
\Bigr({{\rho_{np}}\over{\rho_{pp}}}\Bigl)
\,,
\ee
in terms of the drag coefficient $\rho_{np}$, and the conventional flux quantum, $\Phi_0 ={{hc}\over{2e}}\simeq 2\times 10^{19}\ \rm{G-fm}^2$.\footnote{Note that in the superfluid mixture the flux quantum is {\it not} simply related to fundamental constants; this is a generic feature of two-component superconducting condensates.}

The drag coefficient, $\rho_{np}$, as well as the other superfluid densities, $\rho_{pp}$ and $\rho_{nn}$, depend on the microscopic interactions between the neutron and proton quasiparticles in the interacting mixture, and have been calculated from the BCS theory generalized to a two-component superfluid mixture.~\cite{sau84} For low temperatures, $T \ll \Delta_n , \Delta_p$, these coefficients are given simply in terms of the neutron and proton effective masses,
\ber
\rho_{pp}&=&\rho_p\,\frac{M_p}{M^*_p}\,,
\quad
\rho_{nn} = \rho_n\,\frac{M_n}{M^*_n}\,,
\nonumber
\\
\rho_{pn} &=& \rho_p\,\frac{\delta M_{pn}^*}{M_p^*}\,,
\quad
\rho_{np}  =  \rho_n\,\frac{\delta M_{np}^*}{M_n^*}\,,
\eer
where $\rho_n$ ($\rho_p$) is total the neutron (proton) mass density.
The radial dimension of the flux tube is given by,
\be
\Lambda_* = 29.5\,\left[\frac{M_p^*}{M_p} x^{-1}\rho_{14}^{-1}\right]^{1/2}\,\rm{fm}
\,,
\ee
where $\rho_{14}$ is the mass density in units of $10^{14}{\rm g/cm}^3$ and $x$ is the proton concentration. Typical values of these parameters imply that $\Lambda_* \simeq 50\ \rm{fm}$, and thus the magnitude of the vortex field,
\be\label{B_vortex}
b_{\rm{vortex}} = \frac{|\Phi_*|}{2\pi \Lambda_*^2}
\simeq 3.8\times 10^{15}\,\frac{|\delta M_{np}^*|}{M_p}\,\rho_{14}\,{\rm G}
\,,
\ee
is $b_{\rm{vortex}}\simeq 8\times 10^{14}\ \rm{G}$, which is roughly three orders of magnitude larger than the spin-polarization induced magnetization discussed in Section \ref{Vortices_3P2}.

\section{Electron-Magnetic-Vortex Scattering}\label{Electron_Vortex_Scattering}

The mutual friction timescale resulting from the scattering of the electrons from the magnetic vortices has been calculated in the Born approximation, in Ref.~\onlinecite{sau82a}. The electron Fermi energy for densities of order $10^{14}{\rm g/cm}^3$ is approximately $100\ {\rm MeV}$, which implies that the electrons form an ultra-relativistic degenerate Fermi liquid. In this limit the Born amplitude for electron scattering from the magnetic field of a single vortex is given by
\ber
M(\vk,s \rightarrow \vk',s') 
&=& 
\frac{ec}{2\epsilon_k}\int\frac{d^3x}{\rm V}
e^{i(\vk-\vk')\cdot\vx/\hbar}
\nonumber\\
&\times& 
(\vk+\vk')\cdot\vA(\vx)\delta_{s,s'}
\,,
\eer
where $\vA$ generates the vortex magnetic field given by Eq.(\ref{B_vortex}). The Boltzmann equation for the relaxation of the electron distribution function, $n_{\vk,s}$, following an `instantaneous' change in the relative velocity of the electron fluid and the vortex array is,
\ber
\pder{n_{\vk,s}}{t}
&=& 
N_v\,\frac{2\pi}{\hbar}
\sum_{\vk',s'}|M(\vk,s\rightarrow\vk',s')|^2
\nonumber\\
&\times&\left[n_{\vk',s'}-n_{\vk,s}\right]\,\delta(\epsilon_k-\epsilon_{k'}) 
\,,
\eer
which is simply the total Born scattering rate from $N_v$ vortices calculated from Fermi's Golden rule including the {\it phase space} restrictions imposed by the degenerate sea of electrons. An analysis of this scattering rate shows that the velocity relaxation time between the superfluid core of the star and the plasma is given by~\cite{alp84}
\be
\frac{1}{\tau} = \frac{3\pi^2}{32}\Bigl(\frac{\Omega}{k_{Fe}\Lambda_*}\Bigr)
\Bigl(\frac{\rho_c}{\rho}\Bigr)\left[1-g\Bigl(\frac{\xi}{\Lambda_*}\Bigr)\right]
\,,
\ee
where the dimensionless function, $g(x)$, is given in Ref.~\onlinecite{alp88}, and determines the correction to the electron-vortex scattering due to the finite dimension of the flux line; for $\xi/\Lambda_* \simeq 1$, $g(1) \simeq 0.13$, and is not sensitive to the precise value of the core radius $\xi$ and magnetic field distribution length $\Lambda_*$. At low temperature $(T\ll\Delta_n\,,\Delta_p)$ we find,
\be
\tau\simeq\,400\,\left({{M_p}\over{\delta M_{pn}^*}}\right)^2\,P\,{\rm sec}
\,,
\ee
which implies rapid equilibration of the neutral superfluid interior of the star.

The original two-component model for the dynamical response of a rotating neutron star proposed by Baym et al.~\cite{bay69b} explained the long timescale for the recovery of the period of the Vela and Crab pulsars as a very weak coupling between the neutral liquid core and the crust of the star. In fact the {\it existence} of a neutron superfluid was originally thought to be confirmed by the long timescale for post-glitch relaxation. However, assuming the neutrons and protons are both superconducting then the superfluid drag effect provides an efficient mechanism for the transfer of momentum between the plasma and the neutral superfluid. Equilibration of the core superfluid (actually the establishment of a new steady-state response to the radiation torque) occurs within an hour or so following a glitch. So far the onset of a glitch in either the Crab or Vela has not been observed; typical uncertainties in the onset time of a glitch are a few weeks, although recent glitches in Vela have an uncertainty of one day.~\cite{mcc83} In any event there is as yet no direct observational evidence for a short relaxation timescale, $\tau\sim 10^3\,{\rm sec}\sim 1 {\rm hr}$, involving a major fraction of the moment of inertia of the star. Although Boynton's analysis of the timing noise from Crab (and also from Her X-1) suggests that a large fraction of the moment of inertia of the star is rigidly coupled to the crust, at least on timescales greater than two days.~\cite{boy81}

\section{Superfluidity in the Crust, Vortex Pinning and Glitches}
\label{Glitches}

The {\it origin} of glitches in pulsars is poorly understood. What is clear is that the obvious energy source capable of supplying the enormous energies associated with a glitch, $\Delta E_{rot} = 2\frac{\Delta\Omega}{\Omega} E_{rot}\sim 10^{43} {\rm erg}$, is the rotational energy of the neutron star. However, the physically appealing `starquake' model of the Vela glitches~\cite{rud69,bay71} is unable to account for the magnitude and frequency of the glitches in Vela. It is not possible, based on theoretical estimates of the maximum shear stress that the neutron star crust can sustain, to store $10^{43} {\rm ergs}$ in elastic energy in the crust in a period of 2 to 4 years between glitches.~\cite{bay71,and82}

Metastable states of flow are ubiquitous to superfluids; for example, persistent currents in superfluid helium are a consequence of kinetic energy barriers separating states with different amounts of quantized circulation. That metastability of superflow is a possible explanation for pulsar glitches was suggested by Packard,~\cite{pac72} and a specific model for the source of the metastability was proposed by Anderson and Itoh.~\cite{and75b} This model was motivated by the analogy between the crust of a neutron star and terrestrial {\it hard} superconductors. 
The inner crust $(\rho > 5\times 10^{11}{\rm g/cm^3})$ is a crystalline lattice of heavy nuclei embedded in a degenerate liquid of superfluid neutrons. In the crust the protons are confined within the nuclei, so that unlike the liquid core, there is no superconducting proton component, and since the neutron superfluid in the crust is a condensate of $^1S_0$ pairs there is no electron-magnetic-vortex scattering process present to couple the neutron superfluid to the plasma. However, the existence of the solid crust is expected to have an important effect on the coupling of the neutron superfluid to the crust. In superfluid helium vortices tend to attach themselves to imperfections on the walls of the vessel. If the vessel is decelerated the vortices may remain {\it pinned} to the vessel and a metastable flow is created in which the superfluid is flowing faster than the vessel. Only if the vortices {\it unpin} and annihilate on the vessel wall will the superfluid spin down. A similar metastability exists in laboratory superconductors; very stable current-carrying states of hard superconductors are maintained by the pinning of flux vortices (which are present because of the supercurrent). However, in superconductors pinning occurs on impurities and defects of the crystal lattice. Degradation of the supercurrent occurs only if vortices are transported by the current. In hard superconductors the decay of supercurrents occurs either gradually as vortices {\it diffuse} through the array of pinning sites (vortex `creep') or discontinuously when many vortices unpin and flow unimpeded without re-pinning. These latter events are the laboratory analog of Anderson and Itoh's proposal for the glitch events; as the neutron star slows down the superfluid must expel vortex lines (at a rate of $\dot N = 4M_n \dot\Omega/h\sim 10^9\,{\rm yr^{-1}}$) in order to achieve equilibrium with the crustal rotation. Pinning of this vorticity in the crust is thus a mechanism for storing superfluid kinetic energy. As the relative velocity between the superfluid and crust builds up, the Magnus force tending to expel the vorticity increases, and eventually overcomes the pinning forces. Unpinning occurs at a critical value of the relative angular speed,
\be
|\Omega_s - \Omega_c |_{\rm crit}={{2\pi}\over{\kappa_n}}\ f_p /(R_p\rho_n)\
\,,
\ee
which is determined by balancing the Magnus force per unit length, $f_M = {{\kappa_n}\over{2\pi}} \rho_n |\Omega_c -\Omega_n | \ R_p$, and the pinning force per unit length, $f_p = \epsilon_p/d$, where $\epsilon_p$ is the pinning energy per site, $d$ is the average spacing between pinning centers on a particular vortex and $R_p$ is the radial distance to the pinned vortices. Estimates of the pinning force assume that vortices pin to individual nuclei in the crustal lattice, and that the pinning energy per site is equal to the difference between the vortex core energy in the absence of the pinning center and the condensation energy for neutron pairs bound within the nuclei. The basic equation used to estimate the elementary pinning energy is~\cite{alp77} 
\be\label{Condensation_Pinning}
\epsilon_p=\frac{3}{8}\Bigl[
\rho_i\frac{\Delta(\rho_i)^2}{E_f(\rho_i)} -
\rho_o\frac{\Delta(\rho_o)^2}{E_f(\rho_o)}
\Bigr]\,V
\,,
\ee
where $\rho_i$ and $\rho_o$ are the neutron densities inside and outside the nuclei, and $V$ is the volume of intersection between the vortex core and the nucleus. With energy gaps in neutron matter of order an MeV and plausible assumptions about the number of intersecting nuclei per vortex, the pinning force can be calculated and converted into a critical velocity difference for vortex depinning. Typically, $\delta\Omega_{\rm crit}\sim 10\,{\rm rad/s}$, except perhaps in regions of the inner crust where the pinning force may be an order of magnitude or so smaller.~\cite{alp77,alp84a} In order to account for glitches in terms of vortex depinning every 2-4 years the critical angular velocity difference in some region of the crust must be much smaller than the estimate of 1-10 rad/sec; \ie $\dot\Omega_c\Delta t_{\rm glitch}\simeq 10^{-2}{\rm rad/s}$ (Fig. \ref{fig4}). 

The theory of vortex pinning and flux jumps in laboratory superconductors is not well developed; the elementary pinning energy between a vortex and a small impurity $(R_{\rm imp} \ll \xi)$ was only recently calculated correctly and found to be much larger (by a factor $\xi /R_{\rm imp}$) than the estimate [Eq.(\ref{Condensation_Pinning})] based on minimizing the lost condensation energy of the vortex core and defect.~\cite{thu84} 
Thus, whether regions of weak pinning in neutron stars are likely due to a low density of lattice defects or impurities, or to exceptionally weak intrinsic pinning is unclear. In fact one of the important assumptions made in estimating the pinning energy of vortices in the crust of a neutron star is that vortex lines pin to the nuclear clusters that constitute the crystalline lattice of nuclei. This intrinsic pinning of vortices to the lattice nuclei is not relevant in most superconductors because of the long coherence length compared to the atomic lattice spacing. Superfluidity in the crust may be in the regime where the coherence length is comparable to the size of the nuclei, in which case intrinsic pinning may be relevant; however, there is no microscopic theory of pinning of vortices to the lattice nuclei in short coherence length superconductors. In any event estimates of the vortex pinning energy in the crust are uncertain, but it is difficult to explain the origin and frequency of the Vela pulsar glitches without regions of very weak pinning compared to the estimate of 0.5 MeV/fm. The problem is all the more difficult because within the vortex unpinning model of Ref.~\onlinecite{alp81a} the change in the angular acceleration resulting from the glitch implies that the moment of inertia of the star containing vortices that unpin is $\delta I_p/I_c = \delta\dot\Omega_c/\dot\Omega_c \simeq 10^{-2}$, which translates into roughly $10^{13}$ vortices simultaneously depinning (on any observable timescale) during a glitch. Such a catastrophic unpinning of vorticity is difficult to explain unless there is a mechanism (as yet unspecified) for amplifying fluctuations in the local vortex density which then drive the local superfluid velocity above the critical velocity for unpinning.

In spite of the difficult problem of explaining the {\it trigger} for the catastrophic unpinning of vortex lines, Alpar et al.\cite{alp81b} have analyzed the {\it response} of the crustal superfluid to the glitch (identified as catastrophic unpinning) in terms of the {\it vortex creep} model for vorticity flow,~\cite{and64a} originally invented for understanding the motion of flux in superconductors with defects that pin flux vortices. This model is discussed in detail in this volume by Alpar and Pines; the important point to note here is that vortex creep theory explains the slow relaxation of a pulsar's angular speed back to the pre-glitch spin-down rate in terms of the re-establishment steady-state vortex creep - which requires {\it repinning} of vortex lines. While it seems plausible that this timescale is long, the microscopic physics of the repinning process is not well understood [see Shaham in this volume].

\section{Proton Superconductivity - Some Open Problems}\label{Open_Problems}

Laboratory superconductors exhibit striking properties in response to an applied magnetic field. At sufficiently low magnetic field all supercondutors exhibit the Meissner effect, \ie the complete exclusion of magnetic flux. The threshold field for the penetration of flux into a superconductor depends on the microscopic properties of the superconductor, most importantly the ratio of the field penetration length, $\Lambda = \sqrt{M c^2 / 4 \pi n e^2}$, to the coherence length, $\xi$, which controls the surface energy of a superconducting-normal domain. For $\Lambda/\xi\,>\,\sqrt 2$ (type II superconductors) the surface energy is negative and flux enters the superconductor without destroying the superconducting state in the form of flux lines with an elementary unit of flux, $\Phi_0 = hc/2e$. The threshold field for flux entry is the lower critical field, $B_{c1} = {{hc/2e}\over{\pi \Lambda^2}} ln(\Lambda / \xi )$. Ultimately superconductivity is destroyed when the magnetic field is sufficiently strong, \ie greater than the {\it upper} critical field, $B_{c2} = {{hc/2e}\over{\pi \xi^2}}$. In neutron stars the protons are expected to be type II superconductors with a lower critical field of $B_{c1} \simeq 10^{15} G$ and an upper critical field of roughly $B_{c2} \simeq 10^{17} G$.~\cite{bay69a} Since the stellar field of most neutron stars is estimated to be less than a few times $10^{12} G$, the thermodynamic state of the core of superconducting protons is the Meissner state with complete flux expulsion. However, for a neutron star `born' with a stellar field, \eg in a supernova, the timescale for the flux to diffuse through the high-conductivity ($\sigma$), degenerate plasma may be as long as $\tau_{\rm diffusion}=4\pi\sigma R_{\rm star}^2/c^2 \sim 10^{10}$ years. Therefore, Baym, \et proposed that superconductivity nucleates in the presence of the field by confining the stellar field into a low density of flux tubes, with an average spacing, $d_f\gg\Lambda$.~\cite{bay69a} This implies that the bulk of the neutrons are in a field-free environment in the interior.

\begin{figure}
\centerline{
\includegraphics[width=\columnwidth]{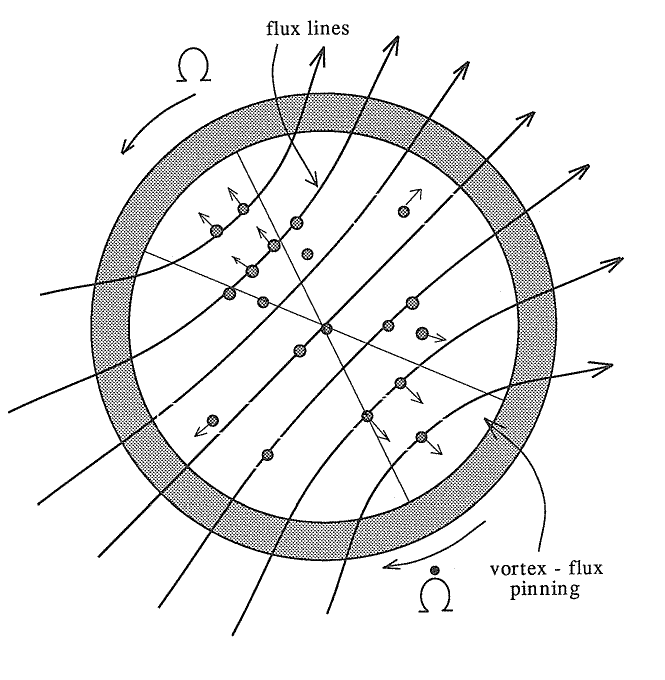}
}
\caption{Vortex lines in the core superfluid may pin on the proton flux lines.  The region of strongest pinning is the cone where the radial flow of vortex lines is nearly perpendicular to the flux lines.}
\label{fig8}
\end{figure}

There are a number interesting unanswered questions regarding the magnetic field structure within the superfluid core. Firstly, there is no detailed theoretical understanding of the non-thermodynamic superconducting transition in the presence of the stellar field, in which the timescale for cooling below $T_c$ is short compared to the flux expulsion timescale $\tau_{\rm diffusion}$. And even if the superconductivity nucleates in the presence of the field, the timescale for the reorganization of the field into quantized flux lines is unknown. Answers to these questions of timescale and flux motion may be relevant to the issue of pulsar `turn off' if indeed the absence of pulsars with apparent ages greater than a few million years old is due to the decay of their magnetic fields. Recently, several authors estimated the Bernoulli and drag forces on proton flux lines and conclude that expulsion of the flux state of the superconductor may occur on the timescale of several million years.~\cite{mus85,jon87} 
However, these authors neglect the tension of flux lines which can act to inhibit flux motion; also the timescale for a flux line to be expelled from the interior is sensitive to the cross-section for electrons scattering off the flux lines.
%

Finally it is interesting to speculate that the proton flux lines (assuming they have nucleated) may have a role in the {\it rotational dynamics} of pulsars. In pulsars the magnetic field axis is misaligned with respect to the rotation axis, so that some of the neutron vortices (which control the rotation of the neutron superfluid) must pass through the proton flux lines as the pulsar spins down. The proton flux lines provide a natural collection of extended pinning `centers' (or rather a `clothesline') for vortex lines in the core of the star; a simple estimate for the pinning energy of a vortex-flux line intersection due to the proton density perturbation in the center of a flux line is
\be
\epsilon_{\rm pin}\sim n \frac{\Delta_p^2}{E_{F_p}^2}\,\frac{\Delta_n^2}{E_{F_n}}
\,(\xi_n^2\xi_p)\lesssim 0.1,{\rm MeV/connection}
\,,
\ee
which suggests that pinning in the superfluid core may be important. In fact there are additional reasons for looking more carefully at the pinning problem in the core superfluid. (i) The effective pinning energy per vortex line is automatically lower than the simple estimate given for pinning in the crust simply because the mean distance between flux lines (pinning centers) is much larger than the distance between the nuclei, $d_f \simeq \Lambda \sqrt{B_{c1}/B} \simeq 10^2-10^3 {\rm fm}$, which translates into a considerably smaller critical velocity difference for unpinning from the flux lines, $\delta \Omega_{crit} \simeq 10^{-2} - 10^{-3} {\rm rad/sec}$, which is reasonably close to the velocity difference that can be built up in $\sim 2$ years as Vela spins down. (ii) Pinning in the crust may be absent or unimportant if intrinsic pinning of vortices to the nuclear lattice is absent (this would be the case if the neutron coherence length overlaps many nuclear clusters) or if the density of crystal defects is low. (ii) Because of the `anisotropy' of the pinning centers in the interior of the star a relatively small {\it cone} of neutron vortices would be pinned by the proton flux lines (see Fig. \ref{fig8}), thus giving rise to a small effective moment of inertia of pinned vorticity, also consistent with the small discontinuity in the spin-down rate due to the glitch. (iii) A model of the post-glitch response based on pinning in the core superfluid, compared to the pinned crustal superfluid, has the advantage of not depending on the difficult problem of vortex {\it repinning} to nuclei in the crust simply because there is no way for vortices flowing radially out to avoid the flux lines in the directions perpendicular to the field. In any case the problem of vortex pinning and dynamics needs additional study in order to determine if catastrophic unpinning and vortex creep are plausible models for pulsar glitches and spin-down of the neutron superfluids.

\section{Acknowledgement}

I benefitted greatly from the lectures and conversations with the participants of the school on {\it Timing Neutron Stars}, and wish to express my thanks to Hakke \"Ogelman, Ed van den Heuvel and the other organizers for their efforts. I particularly thank Ali Alpar for his many stimulating comments to me on the physics of neutron stars over the years. I also thank Daryl Hess and Taku Tokuyasu for their comments on the manuscript. 

%
\end{document}